\documentclass[preprint]{aastex63}

\revised{November 25, 2021} 
\submitjournal{AJ}

\usepackage{lineno}

\shorttitle{IXPE weighted analysis}
\shortauthors{Di Marco et al.}
\DeclareOldFontCommand{\bf}{\normalfont\bfseries}{\mathbf} 
\providecommand{\DIFadd}[1]{{\bf #1}} 
\providecommand{\DIFdel}[1]{} 
\providecommand{\DIFaddbegin}{} 
\providecommand{\DIFaddend}{} 
\providecommand{\DIFdelbegin}{} 
\providecommand{\DIFdelend}{} 
\RequirePackage{listings} 
\RequirePackage{color} 
\lstdefinelanguage{DIFcode}{ 
  moredelim=[il][\color{white}\tiny]{\%DIF\ <\ }, 
  moredelim=[il][\sffamily\bfseries]{\%DIF\ >\ } 
} 
\lstdefinestyle{DIFverbatimstyle}{ 
	language=DIFcode, 
	basicstyle=\ttfamily, 
	columns=fullflexible, 
	keepspaces=true 
} 
\lstnewenvironment{DIFverbatim}{\lstset{style=DIFverbatimstyle}}{} 
\lstnewenvironment{DIFverbatim*}{\lstset{style=DIFverbatimstyle,showspaces=true}}{} 

\begin{document}

\title{A weighted analysis to improve the X-ray polarization sensitivity of IXPE}

\correspondingauthor{Alessandro {Di Marco}}
\email{alessandro.dimarco@inaf.it}

\author[0000-0003-0331-3259]{Alessandro {Di Marco}}
\affiliation{INAF -- IAPS, via Fosso del Cavaliere, 100, Rome, Italy I-00133}

\author[0000-0003-4925-8523]{Enrico Costa}
\affiliation{INAF -- IAPS, via Fosso del Cavaliere, 100, Rome, Italy I-00133}

\author[0000-0003-3331-3794]{Fabio Muleri}
\affiliation{INAF -- IAPS, via Fosso del Cavaliere, 100, Rome, Italy I-00133}

\author[0000-0002-7781-4104]{Paolo Soffitta}
\affiliation{INAF -- IAPS, via Fosso del Cavaliere, 100, Rome, Italy I-00133}

\author[0000-0003-1533-0283]{Sergio Fabiani}
\affiliation{INAF -- IAPS, via Fosso del Cavaliere, 100, Rome, Italy I-00133}

\author[0000-0001-8916-4156]{Fabio {La Monaca}}
\affiliation{INAF -- IAPS, via Fosso del Cavaliere, 100, Rome, Italy I-00133}

\author[0000-0002-9774-0560]{John Rankin}
\affiliation{INAF -- IAPS, via Fosso del Cavaliere, 100, Rome, Italy I-00133}
\affiliation{Università di Roma “La Sapienza”, Dipartimento di Fisica, Piazzale Aldo Moro 2, Rome, Italy I-00185}
\affiliation{Università di Roma "Tor Vergata", Dipartimento di Fisica', Via della Ricerca Scientifica, 1, I-00133 Roma, Italy}

\author[0000-0002-0105-5826]{Fei Xie}
\affiliation{INAF -- IAPS, via Fosso del Cavaliere, 100, Rome, Italy I-00133}

\author[0000-0002-4576-9337]{Matteo Bachetti}
\affiliation{INAF-OAC, Via della Scienza 5, I-09047 Selargius (CA), Italy}

\author[0000-0002-9785-7726]{Luca Baldini}
\affiliation{Università di Pisa, Dipartimento di Ingegneria dell'Informazione, Via G. Caruso 16, I-56122 Pisa, Italy}
\affiliation{INFN-Pisa, Largo B. Pontecorvo 3, I-56127 Pisa, Italy}

\author{Wayne Baumgartner}
\affiliation{NASA Marshall Space Flight Ctr., Huntsville, AL 35812, USA}

\author[0000-0002-2469-7063]{Ronaldo Bellazzini}
\affiliation{INFN-Pisa, Largo B. Pontecorvo 3, I-56127 Pisa, Italy}

\author[0000-0002-9460-1821]{Alessandro Brez}
\affiliation{INFN-Pisa, Largo B. Pontecorvo 3, I-56127 Pisa, Italy}

\author[0000-0003-1111-4292]{Simone Castellano}
\affiliation{INFN-Pisa, Largo B. Pontecorvo 3, I-56127 Pisa, Italy}

\author[0000-0002-3013-6334]{Ettore Del Monte}
\affiliation{INAF -- IAPS, via Fosso del Cavaliere, 100, Rome, Italy I-00133}

\author[0000-0002-7574-1298]{Niccolò Di Lalla}
\affiliation{W.W. Hansen Experimental Physics Laboratory, Kavli Institute for Particle Astrophysics and Cosmology, Department of Physics and SLAC National AcceleratorLaboratory, Stanford University, Stanford, CA 94305, USA}

\author[0000-0003-1074-8605]{Riccardo Ferrazzoli}
\affiliation{INAF -- IAPS, via Fosso del Cavaliere, 100, Rome, Italy I-00133}
\affiliation{Università di Roma “La Sapienza”, Dipartimento di Fisica, Piazzale Aldo Moro 2, Rome, Italy I-00185}
\affiliation{Università di Roma "Tor Vergata", Dipartimento di Fisica', Via della Ricerca Scientifica, 1, I-00133 Roma, Italy}

\author[0000-0002-0984-1856]{Luca Latronico}
\affiliation{INFN-Torino, Via P. Giuria, 1, I-10125 Torino, Italy}

\author[0000-0002-0698-4421]{Simone Maldera}
\affiliation{INFN-Torino, Via P. Giuria, 1, I-10125 Torino, Italy}

\author[0000-0002-0998-4953]{Alberto Manfreda}
\affiliation{INFN-Pisa, Largo B. Pontecorvo 3, I-56127 Pisa, Italy}

\author[0000-0002-1868-8056]{Stephen L. {O'Dell}}
\affiliation{NASA Marshall Space Flight Ctr., Huntsville, AL 35812, USA}

\author[0000-0003-3613-4409]{Matteo Perri}
\affiliation{INAF-OAR, Via Frascati 33, I-00040, Monte Porzio Catone (RM)}

\author[0000-0003-1790-8018]{Melissa Pesce-Rollins}
\affiliation{INFN-Pisa, Largo B. Pontecorvo 3, I-56127 Pisa, Italy}

\author[0000-0002-2734-7835]{Simonetta Puccetti}
\affiliation{ASI, Via del Politecnico snc, I-00133 Roma, Italy}

\author{Brian D. Ramsey}
\affiliation{NASA Marshall Space Flight Ctr., Huntsville, AL 35812, USA}

\author[0000-0003-0411-4243]{Ajay Ratheesh}
\affiliation{INAF -- IAPS, via Fosso del Cavaliere, 100, Rome, Italy I-00133}
\affiliation{Università di Roma “La Sapienza”, Dipartimento di Fisica, Piazzale Aldo Moro 2, Rome, Italy I-00185}
\affiliation{Università di Roma "Tor Vergata", Dipartimento di Fisica', Via della Ricerca Scientifica, 1, I-00133 Roma, Italy}

\author[0000-0001-5676-6214]{Carmelo Sgrò}
\affiliation{INFN-Pisa, Largo B. Pontecorvo 3, I-56127 Pisa, Italy}

\author[0000-0003-0802-3453]{Gloria Spandre}
\affiliation{INFN-Pisa, Largo B. Pontecorvo 3, I-56127 Pisa, Italy}

\author{Allyn F. Tennant}
\affiliation{NASA Marshall Space Flight Ctr., Huntsville, AL 35812, USA}

\author[0000-0001-9194-9534]{Antonino Tobia}
\affiliation{INAF -- IAPS, via Fosso del Cavaliere, 100, Rome, Italy I-00133}

\author[0000-0002-3180-6002]{Alessio Trois}
\affiliation{INAF-OAC, Via della Scienza 5, I-09047 Selargius (CA), Italy}

\author[0000-0002-5270-4240]{Martin C. Weisskopf}
\affiliation{NASA Marshall Space Flight Ctr., Huntsville, AL 35812, USA}

\begin{abstract}

	IXPE is a Small Explorer mission that was launched at the end of 2021 to measure the polarization of X-ray emission from tens of astronomical sources. Its focal plane detectors are based on the Gas Pixel Detector, which measures the polarization by imaging photoelectron tracks in a gas mixture and reconstructing their initial directions. The quality of the single track, and then the capability of correctly determining the original direction of the photoelectron, depends on many factors, e.g., whether the photoelectron is emitted at low or high inclination with respect to the collection plane or the occurrence of a large Coulomb scattering close to the generation point. The reconstruction algorithm used by IXPE to obtain the photoelectron emission direction, also calculates several properties of the shape of the tracks which characterize the process. In this paper we compare several such properties and identify the best one to weight each track on the basis of the reconstruction accuracy. We demonstrate that significant improvement in sensitivity can be achieved with this approach and for this reason it will be the baseline for IXPE data analysis.

\end{abstract}


\section{Introduction} \label{sec:intro}

X-ray astronomy had a substantial break-through with the introduction of X-ray optics mated with imaging detectors in the focal plane. Beside the capability to image extended sources, resolve complex fields and localize with high precision point-like sources, the optics had a dramatic impact on the measurement sensitivity. In experiments with collimators, included the improvements of modulation collimators and coded masks, the photons from the source are compared with the fluctuations of the background counts, measured on the same surface. By means of a telescope the photons collected on the mirrors area, projected on the optical axis, are focused on a focal plane spot of very small surface where a very few background counts are expected. In terms of astrophysics the revolution of the optics, achieved for the first time with HEAO-2/Einstein satellite, enabled imaging of galaxy clusters, supernova remnants and black hole jets and detection of a huge number of extra-galactic sources, introducing X-ray astronomy as a new subject of cosmology. 

Polarimetry was not able to follow this evolution and the main reason was the instrument technology. Any polarimeter is based on a large modulation, namely a high dependence of the response to polarization, but an effective exploitation of the optics in terms of both imaging and background needs also a good localization capability. The only viable polarimeters in the early stage of X-ray astronomy were based on Bragg diffraction at 45$^\circ$ or Compton scattering around 90$^\circ$. Neither technique localizes the interaction point of the impinging photon. The situation evolved with the discovery that, thanks to the new technologies of microelectronics, a finely subdivided gas filled detector could be used as a polarimeter based upon the photoelectric effect. 

The Gas Pixel Detector (GPD) serves as an imaging photoelectric polarimeter, which has been developed in Italy by the Istituto Nazionale di Astrofisica/Istituto di Astrofisica e Planetologia Spaziali (INAF-IAPS) and Istituto Nazionale di Fisica Nucleare (INFN)  over the past $\sim$20 years \citep{Costa2001,Bellazzini2003b,Bellazzini2006,Bellazzini2007,Soffitta21, Sgro}. The Imaging X-ray Polarimetry Explorer (IXPE, \cite{Weisskopf2016,ixpe2,ixpe3,Soffitta}) carries three X-ray telescopes, each with a grazing-incidence mirror assembly and a GPD focal-plane detector \citep{Soffitta, Soffitta21, Sgro} to conduct imaging X-ray polarimetry of cosmic sources. 

In photoelectric polarimeters as the GPD, the $s$ atomic orbital  electrons are ejected from the atom with an azimuthal distribution of cos$^2\phi$, peaked on the direction of the electric vector of the photon. If the azimuthal angles $\phi$ of the initial direction of the photoelectrons were measured perfectly, this polarimeter would have a modulation factor of 100\%. In reality, there are several effects that limit the ideal response of the instrument. Photoelectron path in the gas is traced by the ionization charges produced along the way, which have to drift to the collection plane; they are multiplied and eventually read-out on a pixellated plane with finite pixel size. The quantization of the linear energy loss, the finite pixel dimensions and the transverse diffusion result in an highly blurred image of the track. Moreover, photoelectrons may be scattered close to their generation points, making it difficult to reconstruct their original directions. In the IXPE GPD, photoelectron tracks are analyzed by a custom reconstruction algorithm developed in-house \citep{Bellazzini2003b, Bellazzini03a, Fabiani14, Sgro}, which estimates the initial direction of the photoelectron and the point of the photon absorption. Such an algorithm features several steps: i) a clustering algorithm to distinguish the contiguous physical track from isolated noisy pixels; ii) simple calculations on the moments of the collected charge distribution to distinguish the initial and final part of the track; iii) a refinement of the calculated quantities with a higher weight for the pixels at the beginning of the track. The algorithm calculates also a number of track properties as the track size, the total energy, etc. The purpose of this paper is to explore such parameters to determine which correlate better with the response to linear polarization. These results will be used to weight conveniently the tracks on which a more accurate reconstruction is possible and verify if a better sensitivity can be achieved. 
In the next section the statistical treatment of an event-by-event weighted analysis is presented. In Section 3 a brief explanation of the track reconstruction is given and some different tracks properties are considered to identify the best one to be used as a weight to improve IXPE sensitivity. In section 4 and 5 the weighted analysis is applied to existing IXPE calibration data and to estimate the reachable sensitivity on a reference observation.

\section{Weighted polarization estimation}\label{sec:approach}

A convenient approach to the weighting of data collected by X-ray polarimeters was defined by \cite{Kislat}. The method was originally developed for weighting the response of instruments with non-uniform acceptance, but it can be extended also to increase the contribution of tracks which are better reconstructed because of higher quality. Here we summarize the statistical method presented in \cite{Kislat}. We remark that in this Section and in the following, equations for $\mathcal{U}$ and $\mathcal{Q}$ Stokes parameters are multiplied by a factor 2 with respect to the usual definition. This choice is due to the fact that, as shown in \cite{Kislat}, for photoelectric polarimeters when the usual Stokes parameters definition is applied the polarization degree is $\mathcal{P} = \frac{2}{\mu}\sqrt{Q^2 + U^2}$ while the adopted definition allows to obtain the Equation \ref{eq:pdeg} for $\mathcal{P}$ as expected.

In photoelectric polarimeters, the (normalized) modulation Stokes parameters of a single event are calculated from the the photoelectron emission angle of the k-th event $\phi_k$ and they need to be corrected for the detector response in a later stage analysis 
\begin{eqnarray}
i_k&=&1 \nonumber \\
q_k&=&2\cos2\phi_k \\
u_k&=&2\sin2\phi_k 
  \nonumber
\end{eqnarray}
following an angular distribution \cite{Kislat}:
\begin{equation}
f(\phi) = \frac{1}{2\pi}\left(1+\mathcal{P} \mu \cos (2(\phi-\phi_0)) \right) \label{eq:pdf}
\end{equation}
with $\mathcal{P}$ being the polarization degree, $\phi_0$ the polarization angle and $\mu$ is the modulation factor, that is, the amplitude of the instrumental response to 100\% polarized radiation.

Weights can be introduced, as done in \cite{Kislat}, applying a multiplication factor of the single-event Stokes parameters:
\begin{eqnarray}
i_k&=&w_k \nonumber \\
q_k&=&2w_k\cos2\phi_k\\
u_k&=&2w_k\sin2\phi_k. \nonumber
\end{eqnarray}

The overall Stokes parameters of a measurement of $N$ events are obtained as
\begin{eqnarray}
\mathcal{I}&=&\sum_{k=1}^N w_k \nonumber \\
\mathcal{Q}&=&\sum_{k=1}^N 2w_k\cos2\phi_k\\
\mathcal{U}&=&\sum_{k=1}^N 2w_k\sin2\phi_k .\nonumber
\end{eqnarray}

For the $\phi$ distribution of Eq. \ref{eq:pdf}, the expected $\mathcal{U}$ and $\mathcal{Q}$ values can be estimated
\begin{eqnarray}
\langle \mathcal{Q} \rangle &=& \sum_{k=1}^N 2 w_k \int_0^{2\pi} \cos 2 \phi f(\phi) d\phi = \mathcal{I} \mathcal{P} \mu \cos(2\phi_0) \label{eq:expected}\\
\langle \mathcal{U} \rangle &=& \sum_{k=1}^N 2 w_k \int_0^{2\pi} \sin 2 \phi f(\phi) d\phi = \mathcal{I} \mathcal{P} \mu \sin(2\phi_0)
\nonumber
\end{eqnarray}
then the polarization degree is
\begin{equation}
\mathcal{P}=\frac{\sqrt{\mathcal{U}^2+\mathcal{Q}^2}}{\mu\mathcal{I}}=\frac{\sqrt{Q^2+U^2}}{\mu}\DIFaddbegin \label{eq:pdeg}\DIFaddend ,
\end{equation}
where $Q$ and $U$ are the $\mathcal{Q}$ and $\mathcal{U}$ Stokes parameters normalized by $\mathcal{I}$ and $\mu$ is the modulation factor, that is, the amplitude of the instrumental response to 100\% polarized radiation. 

Following a similar approach, the expected variances for the Stokes parameters can be determined:
\begin{eqnarray}
Var(\mathcal{Q}) &=& 
W_2 \int_0^{2\pi} \left(2 \cos 2\phi - \mathcal{P} \mu \cos (2\phi_0) \right)^2 f(\phi) d\phi = \nonumber \\
&=&W_2 \left[ 2-\mu^2 \mathcal{P}^2 \cos^2(2\phi_0)  \right] \label{eq:var} \\
Var(\mathcal{U})& = & W_2 \left[ 2-\mu^2 \mathcal{P}^2 \sin^2(2\phi_0)
 \right]  \nonumber
\end{eqnarray} 
where we introduced the following quantity:
\begin{equation}
W_2=\sum_{k=1}^N w_k^2.
\end{equation}
From the previous equations, we can determine the reconstructed Stokes parameters to compare with results from theory or other experimental values as $Q_r = Q/\mu$ and $U_r = U/\mu$. For these latter values we can obtain the following uncertainties in case of a large data set
\begin{eqnarray}
\sigma_{Q_r}&=&\sqrt{\frac{W_2}{\mathcal{I}^2}\left( \frac{2}{\mu^2}-Q_r^2\right)} \label{eq:unc_stokes}\\
\sigma_{U_r}&=& \sqrt{\frac{W_2}{\mathcal{I}^2}\left( \frac{2}{\mu^2}-U_r^2\right)}. \nonumber
\end{eqnarray}
Following the same approach the covariance of $\mathcal{Q}$ and $\mathcal{U}$ can be obtained:
\begin{eqnarray}
Cov(\mathcal{Q}, \mathcal{U})&=& -\frac{W_2}{2} \mathcal{P}^2 \mu^2 \sin(4\phi_0)
\end{eqnarray}
and from this latter one we obtain
\begin{equation}
Cov(Q_r,U_r) = - \frac{W_2}{2\mathcal{I}^2} \mathcal{P}_r^2 \sin(4\phi_r)
\end{equation}
and the Pearson correlation coefficient is
\begin{equation}
\rho(Q_r,U_r) = -\frac{\mathcal{P}_r^2\sin(4\phi_r)}{\sqrt{16 - 8 \mathcal{P}_r^2 + \mathcal{P}_r^4\sin^2(2\phi_r)}}
\end{equation}
From these latter relations it is possible to observe that a correlation between U and Q is present for higher values of polarization degree and $\mu$, in case of low polarization degree this correlation goes to zero.

In \cite{Kislat} a joint error posterior distribution for the polarization degree and angle estimators has been derived to include the correlation between the Stokes parameters
\begin{eqnarray}
P(\hat{\mathcal{P}},\hat{\phi_0}|\mathcal{P},\phi_0)& =& \frac{\sqrt{N_{eff}}\hat{\mathcal{P}}^2\mu^2}{2\pi\sigma} \times \nonumber \\ & & exp\left(-\frac{\mu^2}{4\sigma^2} \left[ \hat{\mathcal{P}}^2+\mathcal{P}^2 - 2\hat{\mathcal{P}}\mathcal{P}\cos(2(\phi_0-\hat{\phi_0})) - \frac{\hat{\mathcal{P}}^2\mathcal{P}^2\mu^2}{2}\sin^2(2(\phi_0-\hat{\phi_0})) \right] \right)
\end{eqnarray}
with $\hat{\mathcal{P}}$ and $\hat{\phi_0}$ obtained values of polarization degree and angle respectively and
\begin{equation}
\sigma=\sqrt{\frac{1}{N_{eff}}\left( 1 - \frac{\mathcal{P}^2\mu^2}{2}\right)},
\end{equation}
where we define the ``effective'' number of counts of the measurement $N_\textrm{eff}=\mathcal{I}^2/W_2$. 

This distribution, when the correlation term is negligible, can be reduced to the Rice distribution as shown by \cite{Vaillancourt2006}. In cases with $\mathcal{P}$ and $\mu$ not close to 0, the Gaussian approximation for the marginalized errors can be used as stated in \cite{Kislat}: 
\begin{equation}
\sigma^2_\mathcal{P} \simeq \frac{2-\mathcal{P}^2\mu^2}{(N_{eff}-1)\mu^2}.
\label{eq:mod_std}
\end{equation}
in case the correlation term can be neglected, it reduces after marginalization to
\begin{equation}
\sigma_\mathcal{P}^2 \simeq \frac{2}{\mu^2N_{eff}}.
\end{equation}
From this equation it is possible to see that minimizing $\sigma_\mathcal{P}$ corresponds to maximizing the modulation factor.

The Minimum Detectable Polarization, MDP$_{99}$, quantifies the polarization sensitivity at 99\% confidence level. For the weighted analysis it is
\begin{equation}
\textrm{MDP}_{99}\approx \frac{4.29}{\mu_{weighted}} \sqrt{\frac{W_2}{\mathcal{I}^2}}=\frac{4.29}{\mu_{weighted}\sqrt{N_{\textrm{eff}}}},
\end{equation}
in analogy of the corresponding expression for the unweighted analysis:
\begin{equation}
\textrm{MDP}_{99,\textrm{unweighted}}\approx\DIFdelbegin \DIFdel{\frac{4.29}{\mu\sqrt{N}}}\DIFdelend \DIFaddbegin \DIFadd{\frac{4.29}{\mu_{unweighted}\sqrt{N}}}\DIFaddend .
\end{equation}

An older, alternative analysis method is based on a cut analysis, the so-called ``standard cuts'' \citep{Muleri16}. This one is carried out applying a two step selection of the events where 20\% of them are removed. The first group of events is removed applying an energy cut: the energy spectrum is fitted with a Gaussian and events outside $\pm 3 \sigma$ from the peak center are removed. The events surviving the energy selection are ordered as a function of their ``eccentricity'' and lower-eccentricity tracks are removed up to a threshold which removes 20\% of the initial events including the ones removed from the energy selection. 

This method and every kind of cut/selection analysis is similar to a weighted analysis, by the fact that data selection is a simple kind of weights where good events have $w_k=1$ and the bad ones $w_k=0$. In this case, the effective number of events is equal to the number of the good ones, while bad ones are removed by the analysis. 

``Standard cuts'' were developed for the analysis of monochromatic laboratory sources and they cannot be applied to observations of astrophysical sources with continuum spectra, which will be carried out with IXPE. As an alternative approach is needed to obtain a suitable sensitivity, we propose this weighted analysis. It is worth noting that, when weights are introduced, $N_\textrm{eff}<N$ such that a better sensitivity is achieved only if the decrease in the square root of counts' effective number is balanced by a sufficient increase of modulation factor.

\section{Determination of event-by-event best-weight parameter}

\begin{figure}[h]
	\centering
	\includegraphics[width=9cm]{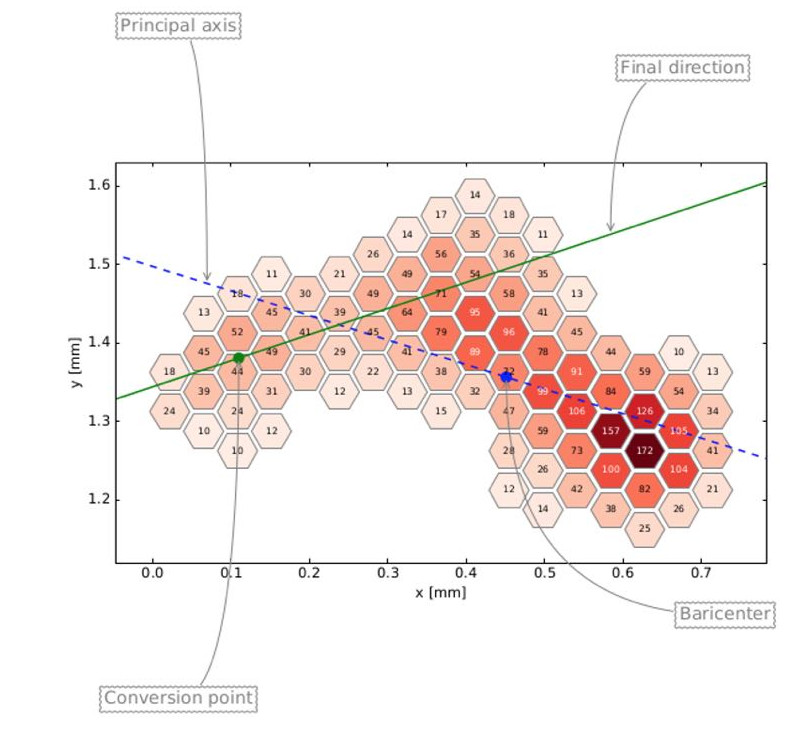}
	\caption{Example of ionization track resulting from the absorption of a 5.9-keV X-ray in the GPD. The reconstructed photoelectron direction and impact point are shown \citep{Sgro17}.
	}
	\label{fig2}
\end{figure}

Photoelectron tracks collected by the GPD on-board IXPE are analyzed by a custom algorithm to extract relevant information. In this algorithm the photoelectron track direction is determined on the basis of a two-step moment analysis that has been refined over the years \citep{Bellazzini2003b,Bellazzini03a,Fabiani14,Sgro}. In the following we summarize such an algorithm to provide context for the following analysis. In the first step the barycenter of the charge distribution (as the one of Figure \ref{fig2}) is calculated by
\begin{eqnarray}
x_B = \frac{\sum_i q_i x_i}{\sum_i q_i} \\ \nonumber
y_B = \frac{\sum_i q_i y_i}{\sum_i q_i},
\label{eq:bar}
\end{eqnarray}
where $x_i$ and $y_i$ are the coordinates of pixels in the image and $q_i$ is the charge collected in it. 

Centering the track in $(x_B, y_B)$, the second moment of the charge distribution is
\begin{equation}
M_2(\psi) = \frac{\sum_i q_i \left[ (x_i - x_B) \cos (\psi) + (y_i - y_B) \sin (\psi) \right]^2}{\sum_i q_i},
\end{equation}
where the angle $\psi$ is between the track direction and the $x$ axis. This is used to determine the axis for the charge distribution of minimal or maximal extension: in fact, the angle which maximizes or minimizes $M_2$ is obtained imposing $dM / d\psi = 0$. The maximum and minimum angles ($\psi_{max}$ and $\psi_{min}$ respectively) are 90$^\circ$ apart and the allow to determine the longitudinal and transverse second moment (corresponding to Track Length, TL, and Track Width, TW, respectively). At this point $\psi_{max}$ defines the angle between the track direction and the $x$-axis. The third moment estimated for $\psi_{max}$
\begin{equation}
M_3(\psi_{max}) =  \frac{\sum_i q_i \left[ (x_i - x_B) \cos (\psi_{max}) + (y_i - y_B) \sin (\psi_{max}) \right]^3}{\sum_i q_i},
\end{equation}
is used to determine the X-ray impact point position with respect to the $(x_B, y_B)$ coordinates. In fact, this is the less dense part, as the photoelectron loses more and more energy as it slows down, forming the so-called Bragg peak. In Figure \ref{fig2} it is possible to observe that the Bragg peak region is more dense than the initial part of the track: however the initial photoelectron emission direction in this part of the track is lost because of the photoelectron scattering within the gas cell. 

In the second step the distance between each pixel and the $(x_B, y_B)$ coordinates is estimated in $M_2$ units, pixels that have a sign of the distance different with respect to the one of the $M_3(\psi_{max})$ are removed, pixels having a distance value within a horseshoe region centered at $(x_B, y_B)$ with a minimum and maximum radius defined by 1.5 and 3.5, respectively, are selected and used to determine the impact point coordinates, as explained in the following. This method allows to exclude pixels due to the Bragg peak or Auger electrons. To take into account the energy loss in the gas cell, pixel charges  are weighted by $w=e^{-\frac{D}{w_0}}$, with $D$ distance from the impact point and $w_0 = 0.05$. The weighted charges in each pixel are then used to estimate the $M_2$ in the horseshoe region and to obtain the the final photoelectron track direction.

\begin{figure}[h]
	\centering
	\includegraphics[width=0.45\textwidth]{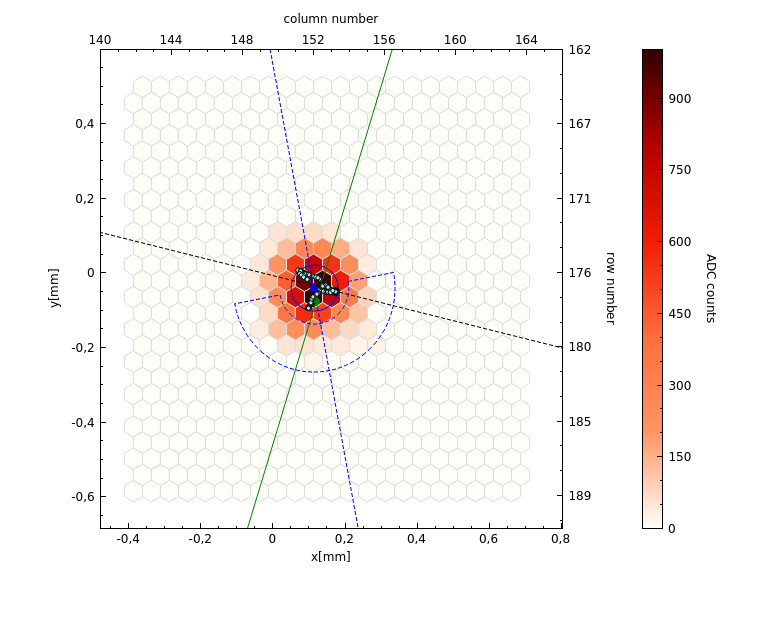}
	\includegraphics[width=0.45\textwidth]{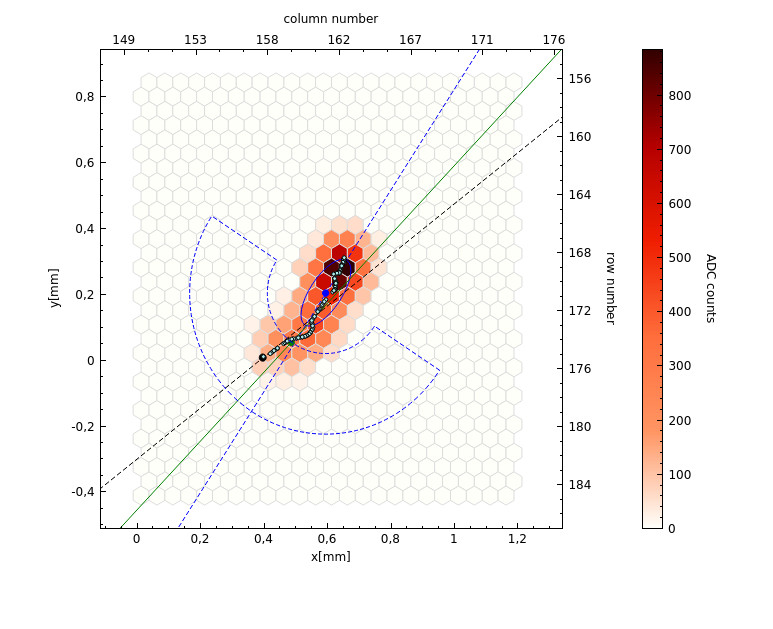}
	\caption{Two examples of simulated tracks at about 4 keV from a Gas Pixel Detector: $Left$ a circular track with $\alpha \simeq 0.01$; $Right$ a quasi--linear track with $\alpha \simeq 0.7$. In these plots, the green line represents the Monte Carlo true direction; the dotted-blue line, the first-step reconstructed direction; and the dotted-black line, the final reconstructed direction after Bragg removal (see text).}\label{fig:tracks}
\end{figure}

In Figure \ref{fig:tracks} we report the horseshoe region, the direction determined in the first and the second step, but also the true one obtained from a Monte Carlo simulation. It is possible to observe that in cases where the initial part can be well identified with respect to the Bragg peak region, the initial direction of the photoelectron can be reconstructed with a better accuracy, as the tracks are of better quality (Figure \ref{fig:tracks}-$right$). In  case, as in Figure \ref{fig:tracks}-$left$, where the Bragg peak is not well identified, the track direction is not well determined.

We compared several quantities calculated by the GPD track-analysis algorithm to obtain an estimator of the track quality, to be used as a weight in the subsequent analysis. In the following we present only some of them which are related to the topological properties of the tracks, in particular:
\begin{itemize}
	\item Track Length (TL): longitudinal second moment of the track;
	\item Track Width (TW): transverse second moment of the track;
	\item Track Size (TS): number of pixel over threshold in the track;
	\item Track Elongation (TE): ratio of Track Length and Track Width;
	\item Ellipticity, defined as
	\begin{equation}
	\alpha = \frac{TL-TW}{TL+TW},
	\end{equation} which quantifies the track's elongation. It is defined in the interval [0;1], being 0 for circular tracks and 1 for linear tracks;
	\item Skewness of the track, defined by the third moment.
\end{itemize}

To relate these parameters with the response to polarization, we used the IXPESIM Monte Carlo tool developed by the IXPE team \citep{Manfreda} and based on Geant4 to simulate a large data set. In particular, we generated $\simeq 10^7$ events produced by 100\% polarized source and distributed in energy with a power-law spectrum with photon index $-2$. 
To obtain an approximate but quick evaluation, we assumed that the best quantity is the one that provides the smallest uncertainty on the degree of polarization, defined in the previous section, that means to maximize the modulation factor. 

In \cite{Marshall2021} it has also been demonstrated that in weighted analysis, the modulation factor itself is the best weight parameter. This means that we need to find in the following the parameter which allows to minimize the $\sigma_{\mathcal{P}}$ and that is also a good proxy for the modulation factor. We notice that we could have used also the uncertainty on the angle (as in Equation 37 of \citet{Kislat}) but our choice is  consistent with the use of the MDP$_{99}$ (and not on the error on the angle) as the primary factor of merit of a polarimeter. Moreover both approaches converge, in practice, to the selection of the modulation factor as the driver.

\begin{figure}
	\begin{center}
		\begin{tabular}{c}
			\includegraphics[height=0.155\textheight]{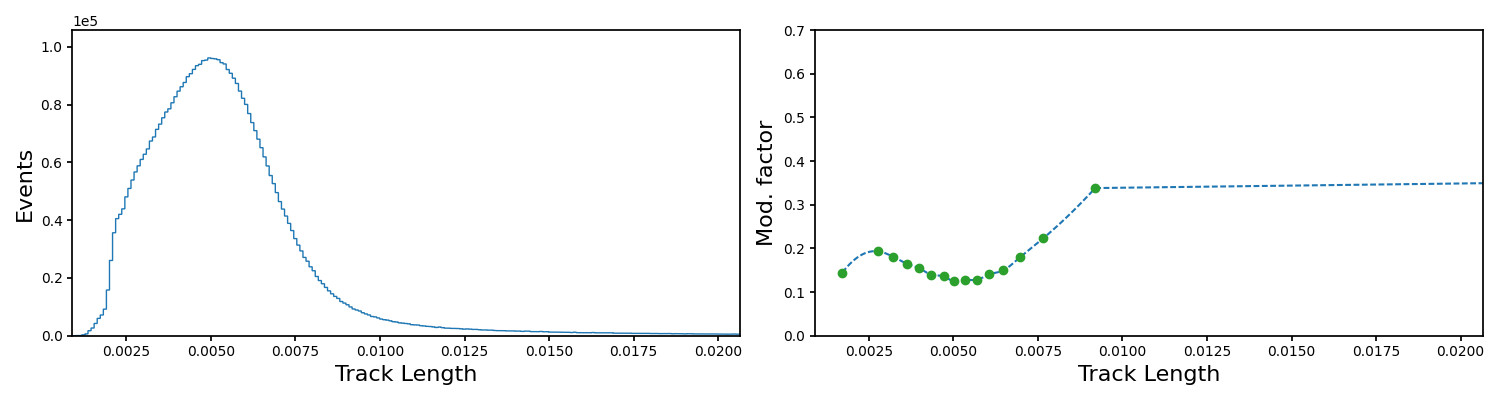} \\
			\includegraphics[height=0.155\textheight]{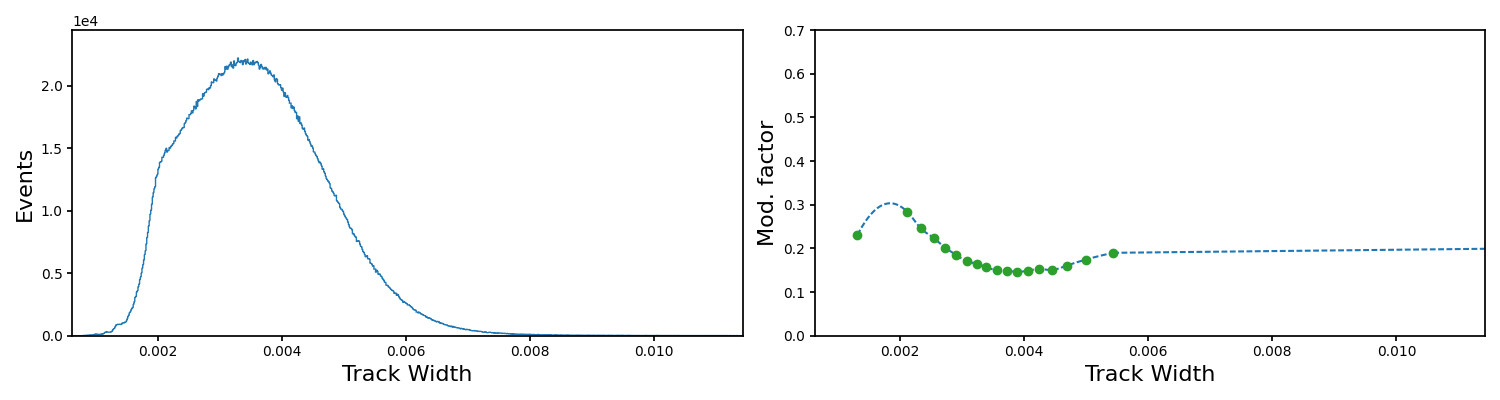} \\	
			\includegraphics[height=0.155\textheight]{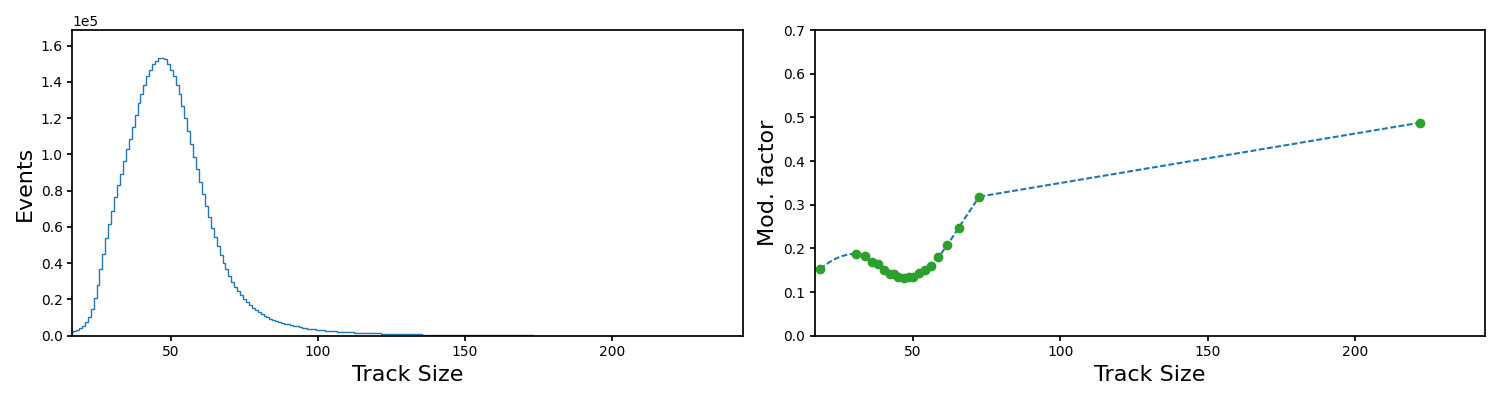} \\		
			\includegraphics[height=0.155\textheight]{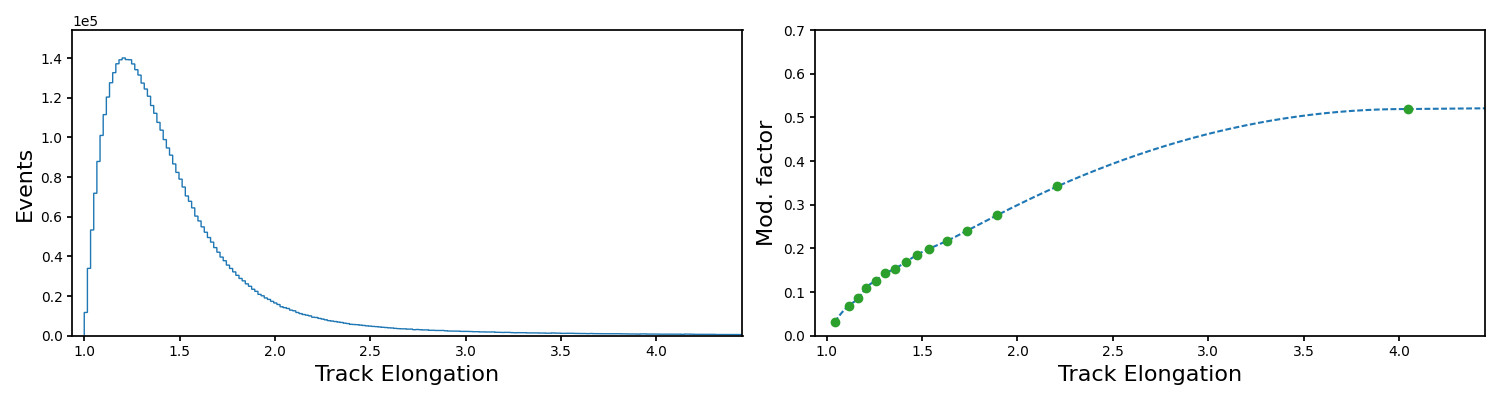} \\	
			\includegraphics[height=0.155\textheight]{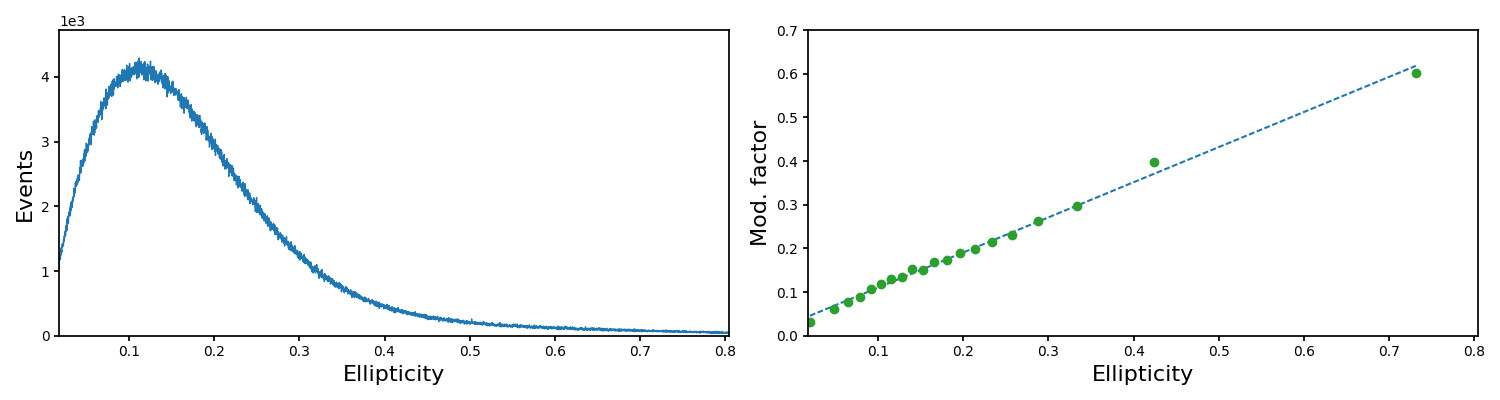} \\
			\includegraphics[height=0.155\textheight]{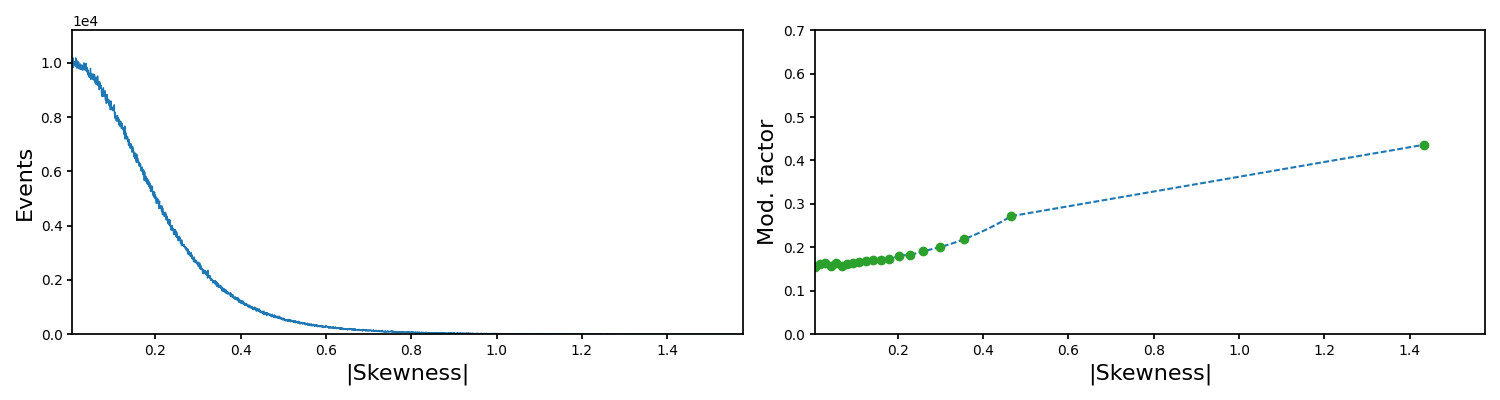}		
		\end{tabular}
	\end{center}
	\caption 
	{ \label{fig:sim_alpha} Distribution of event's TL, TW, TS, TE, Ellipticity and Skewness absolute value for simulated data (left) and modulation factor as a function of the considered parameter (right).} 
\end{figure}

The distributions of the parameters under study are shown in Figure \ref{fig:sim_alpha}. The data have been divided in 20 bins with the same number of events. For each bin, the modulation factor $\mu_j$ is calculated, as shown in Figure \ref{fig:sim_alpha} and we estimated the $\sigma^2$ on the whole data set as $\sigma^2_\mu = \frac{1}{\sum_{j=1}^{20} \frac{1}{\sigma_{\mu_j}^2}}$; $\sigma^2$ values are reported in Table~\ref{tab:sim_sigma}.
From this analysis it is evident that the Track Ellipticity provides the lowest $\sigma_{\mathcal{P}}$ value and is also linearly related with the modulation factor.
\begin{table}[b]
	\caption{Modulation factor variance for each of the parameter considered as a weight. The data set is obtained by Monte Carlo simulations for a power-law energy spectrum in the energy range 2--8 keV and 500000 events.} 
	\label{tab:sim_sigma}
	\begin{center}    
		\begin{tabular}{|r|c|}
			\hline\hline
			\textbf{Parameter} & \textbf{$\sigma^2 (\times 10^{-5})$} \\ \hline\hline
			Track Length			& 1.98 \\
			Track Width     		& 4.40 \\
			Track Size       		& 2.59 \\
			Track Elongation 		& 2.02 \\
			Track Ellipticity 		& 1.90 \\
			Abs. Track Skewness		& 4.23 \\
			\hline 
		\end{tabular}
	\end{center}
\end{table} 
The results obtained above with Monte Carlo simulations are confirmed with a representative measurement carried out during the calibration of the IXPE focal-plane detector with a 3.69~keV polarized source, described in \cite{Muleri2021} as shown by the results in Table~\ref{tab:data_sigma}. The numerical values of Table~\ref{tab:sim_sigma} and Table~\ref{tab:data_sigma} refer to different data sets with a different energy spectrum and a different number of events, but they show the same best-weight parameter --- namely, $\alpha$.
\begin{table}[ht]
	\caption{Modulation Factor variance for each parameter considered as a weight for ground calibration data measured at 3.69~keV.} 
	\label{tab:data_sigma}
	\begin{center}       
		\begin{tabular}{|r|c|}
			\hline\hline
			\textbf{Parameter} & \textbf{$\sigma^2 (\times 10^{-5})$} \\ \hline\hline
			Track Length			& 3.06 \\
			Track Width     		& 3.33 \\
			Track Size       		& 3.51 \\
			Track Elongation 		& 2.78 \\
			Track Ellipticity 		& 2.77 \\
			Abs. Track Skewness		& 3.00 \\
			\hline 
		\end{tabular}
	\end{center}
\end{table} 
Both elongation and ellipticity provide good performance, but we choose the latter because its interval is conveniently between 0 and 1 and the modulation factor grows linearly with the ellipticity parameter following the function $\mu = 0.03 + 0.804 \alpha$.

\begin{figure}[b]
	\begin{center}
		\begin{tabular}{c}
			\includegraphics[width=0.5\textwidth]{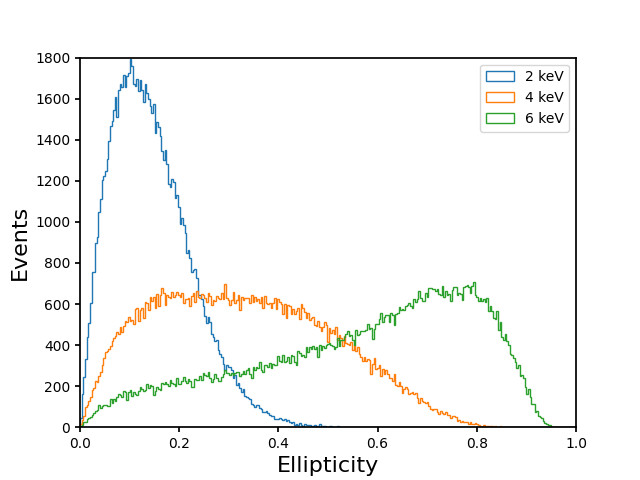}
			\end{tabular}
	\end{center}
	\caption 
	{ \label{fig:alpha_ener} Distribution of Ellipticity, $\alpha$, at 2 keV (blue line), 4 keV (orange line), 6 keV (green line).} 
\end{figure}

It is worth noting that $\alpha$ is not a mere proxy of the energy. In Figure \ref{fig:alpha_ener} the $\alpha$ distributions at 2 keV, 4 keV and 6 keV have been estimated by Monte Carlo simulations. One can see that, for a monochromatic source, $\alpha$ is distributed over a wide range of values.

\begin{figure}[!h]
	\begin{center}
		\begin{tabular}{c}
			\includegraphics[width=0.4\textwidth]{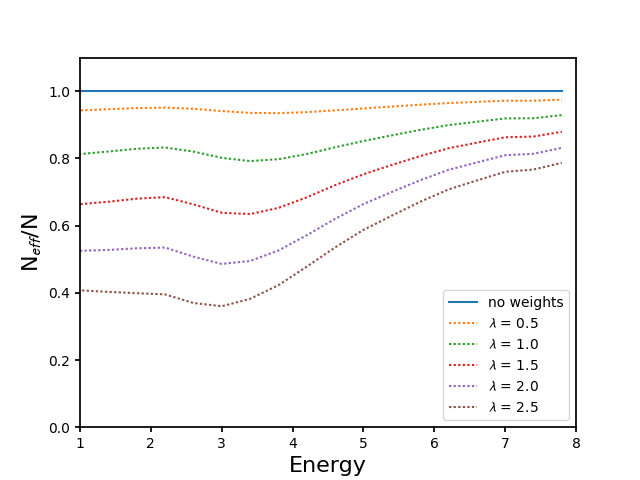}
			\includegraphics[width=0.4\textwidth]{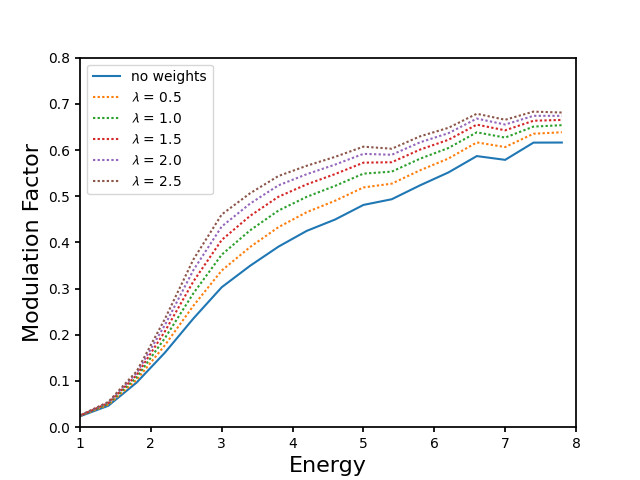}\		\end{tabular}
	\end{center}
	\caption 
	{ \label{fig:lambda_choice1} $N_{eff}/\mathcal{I}$ ($left$) and modulation factor ($right$) as a function of the energy at different $\lambda$ values, as obtained by Monte Carlo simulations. See the text.} 
\end{figure}

\begin{figure}[!hb]
	\begin{center}
		\begin{tabular}{c}
			\includegraphics[width=0.4\textwidth]{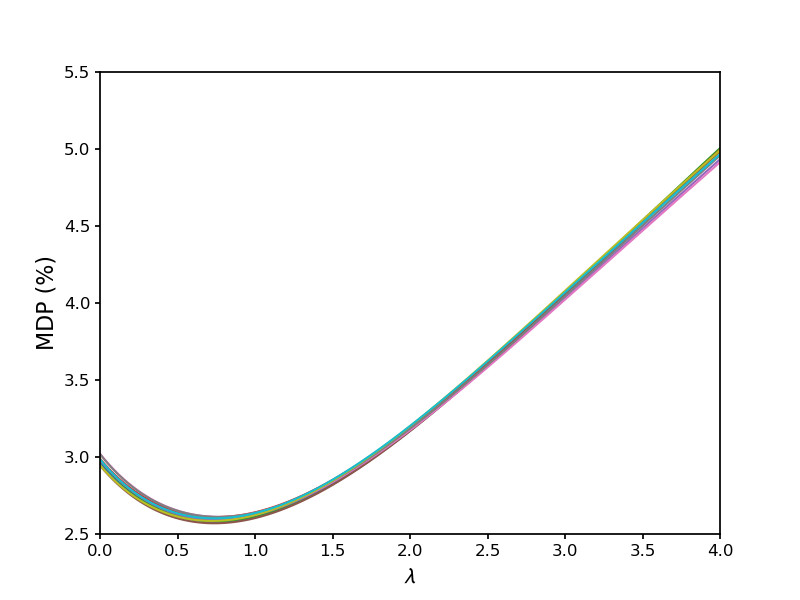}
		\end{tabular}
	\end{center}
	\caption 
	{\label{fig:lambda_choice2} MDP$_{99}$ values as a function of $\lambda$ for 10 different Monte Carlo simulations. All these distributions have a minimum around $\lambda=0.75$} 
\end{figure}

Being that ellipticity is the best parameter for the weighted analysis, we tried to further improve the sensitivity using as a weight some function of $\alpha$ --- namely, $\alpha^\lambda$. To choose the best value of $\lambda$, we applied it to the simulated data set and calculated for different values of $\lambda$ the following quantities:
\begin{itemize}
	\item modulation factor as a function of the energy;
	\item reduction of the effective-count fraction, $N_{eff}/\mathcal{I}$, as a function of the energy;
	\item MDP$_{99}$ calculated as the mean value of 10 Monte Carlo simulations. From this latter plot we conclude that MDP$_{99}$ has a minimum for $\lambda \simeq 0.75$;
\end{itemize} 
The results are reported in Figure \ref{fig:lambda_choice1} and \ref{fig:lambda_choice2}. As we increase the value of $\lambda$, the weighting of data is more and more important with the effect that the number of effective counts decreases but the modulation factor increases. When we combine the two to obtain the sensitivity expressed with the MDP$_{99}$, the best value for sensitivity -- that is the minimum in MDP$_{99}$ -- is obtained for $\lambda=0.75$.

\section{Application of the weighted analysis to IXPE ground calibration data}\label{performances}

In this section we applied the weighting approach to calibration data obtained with IXPE Detector Units (DUs) during ground calibrations \citep{calibration}. As results are consistent for all the IXPE DUs, we present, as a representative case, results for DU-FM2.

\begin{figure}[!h]
	\begin{center}
		\begin{tabular}{c c c}
			(a) & (b) & (c) \\
			\includegraphics[width=0.3\textwidth]{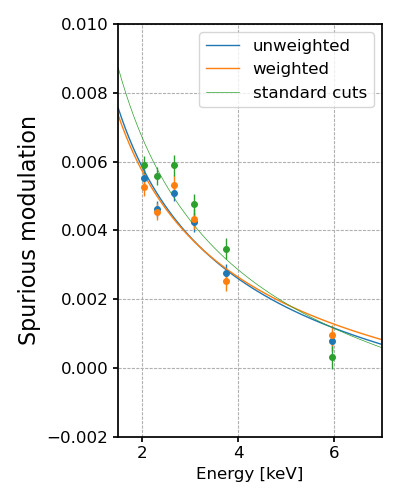} &
			\includegraphics[width=0.3\textwidth]{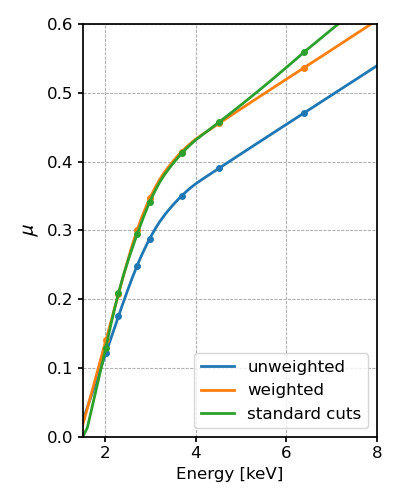} &
			\includegraphics[width=0.3\textwidth]{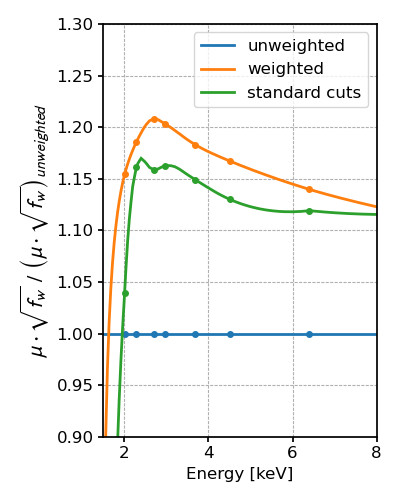} 
		\end{tabular}
	\end{center}
	\caption 
	{\label{fig:comparison} Comparison as a function of energy for three different analyses (unweighted, weighted and ``standard cuts'') displaying: (a) measured spurious modulation; (b) measured modulation factor; (c) relative quality factor. Although values are shown for the representative case of IXPE DU-FM2, similar behaviors are obtained for the other DUs aboard IXPE.} 
\end{figure}

Data analysis has been performed for both unpolarized and polarized sources to derive the response to completely polarized and unpolarized radiation. In particular, in this section we derive the modulation factor and the spurious modulation amplitude --- that is the detector response to unpolarized radiation --- by using the following analyses:
\begin{itemize}
	\item unweighted/uncut: Analysis without cuts or weights, where all events are included in the analysis; 
	\item weighted/uncut: Analysis without cuts but with an event-by-event weight equal to $\alpha^{0.75}$.
	\item std\_cut/unweighted: Analysis with ``standard cuts'' to remove 20\% of the data but without weights.
\end{itemize}
Figure \ref{fig:comparison}a shows the spurious modulation as a function of energy as measured (dots). Note that the unweighted and weighted analysis give similar results, smaller than the ones obtained with ``standard cuts''. This component is removed in the analysis with an algorithm described by \cite {Rankin21}. 
The modulation factor as a function of the energy is reported in Figure \ref{fig:comparison}b. Modulation factor with weights and ``standard cuts'' is higher than the value obtained with the unweighted analysis. 

The guideline for these analysis approaches is always the maximization of the so-called ``Factor of Quality'' \citep{Pacciani2003}. This is the product of the modulation factor by the square root of the efficiency, which had been already used effectively for the selection of the design parameters of the detectors. Thus, to have an overall idea of the improvement in sensitivity, taking into account the $N_{eff}$ contribution with experimental data, the quality factor is given by $\mu \sqrt{f_{w}}$, where $f_w$ is the fraction of events $N_{eff}/N$ after cuts/weights analysis. In Figure \ref{fig:comparison}c the quality factor for each analysis is normalized to the unweighted case, showing that the weighted analysis provides the best improvement in sensitivity over the entire IXPE energy range.

\section{Sensitivity improvement on a simulated IXPE observation}
\label{sec:sensitivity}
\begin{figure}[b]
	\begin{center}
		\begin{tabular}{c}
			\includegraphics[width=0.5\textwidth]{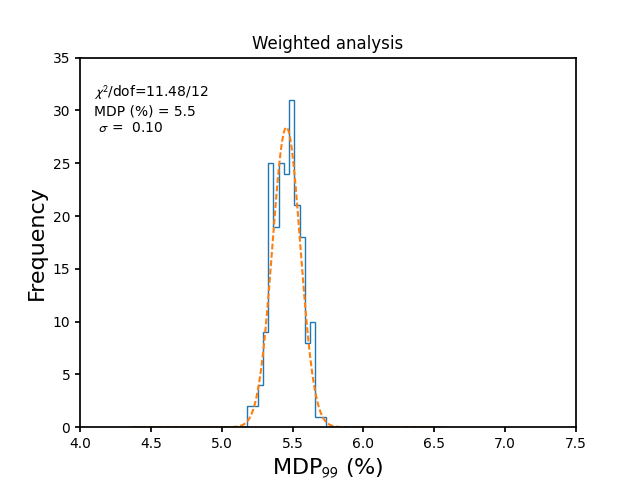}
			\includegraphics[width=0.5\textwidth]{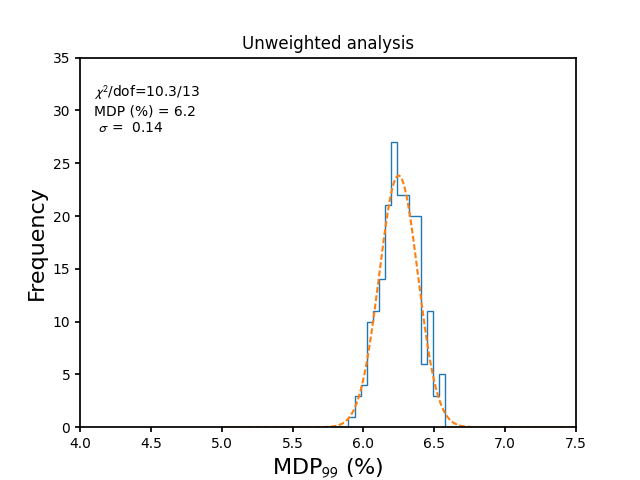}

		\end{tabular}
	\end{center}
	\caption 
	{ \label{fig:mdp_dist} Distribution of MDP$_{99}$ values for 200 IXPE simulated observations of a point-like source with energy spectrum described by a power-law with number index -2, a flux 10$^{-11}$ erg/cm$^2$/s in the energy range 2--8 keV and 10 days of integration time. MDP$_{99}$ values are estimated for weighted ($Left$) and unweighted ($Right$) analysis. See text.}
\end{figure}

Above we evaluated the improvement achieved by applying the weighted analysis to a monochromatic beam, albeit for different energies. However IXPE (and any other polarimetry mission except the Bragg-diffraction polarimeters) integrates photons within a continuous energy band, in order to achieve adequate sensitivity. Depending upon the celestial source modeling or its intensity, it could be the whole energy range of the instrument. We want to know how the improvements detected at single energy values convert into an improved detection capability for a continuum source.

To estimate the improvement of the weighted analysis with respect to the unweighted one, we simulated an observation with IXPE of a reference point-like source with a power-law spectrum (number index -2), a flux of 10$^{-11}$ erg/cm$^2$/s in the energy range 2--8 keV and 10 days of integration time. In the simulation, we considered the GPD quantum efficiency at launch (following the model described by \cite{Sgro} and applying to each IXPE DU the expected DME pressure at launch), the UV filters transparency and the optics effective area as measured at NASA Marshall Space Flight Center during ground calibrations of the optics. We simulated 200 data sets in IXPESIM obtaining MDP$_{99}$ values in the range (5.2--5.7)\% for the weighted analysis and (5.9-6.6)\% for the unweighted one. Figure \ref{fig:mdp_dist} shows the distribution of these values, which are compatible with a Gaussian distribution with mean values 5.5\% and 6.2\% for the weighted and the unweighted analysis, respectively.

This kind of analysis cannot be performed with the ``standard cuts'' approach because the energy spectrum of the source is continuous. This result shows that the weighted analysis by using $\alpha^{0.75}$ as an optimal weight can allow a significant improvements in polarization sensitivity. The weighted analysis achieves a sensitivity within the IXPE scientific requirement that MDP$_{99}$ not exceed 5.5\% for the reference source.

\section{Conclusions}

In this paper we applied the weighted analysis described in \cite{Kislat} to data collected by the focal-plane X-ray polarimeter on-board IXPE, which are the GPDs. Starting from  quantities calculated using the dedicated track reconstruction analysis developed for the GPD, we identified a best-weight parameter for both simulated and real data. The weighted analysis improves the IXPE sensitivity.

In this analysis we showed that the modulation factor is affected significantly by weights and its value depends upon the ``strength'' of the weighting. However, the gain on sensitivity, expressed as a reduction of MDP$_{99}$, is partially offset by a reduction in the effective number of events ($N_{eff}$). Among different choices for the weighting, we used the track ellipticity, which provides the best sensitivity and has the advantage that its range is [0,1].  

In such an assumption, when the analysis is applied to calibration data we conclude the following:
\begin{itemize}
	\item Application of weights reduces spurious modulation with respect to values obtained with ``standard cuts'' used in IXPE calibration analysis;
	\item Modulation factor is larger than the value obtained with ``standard cuts'' or unweighted/uncut analysis.
\end{itemize}

We evaluated by means of Monte Carlo simulations the improvement in sensitivity considering a Crab-like spectrum source. Our analysis showed that, with respect to an unweighted scheme, the weighted scheme improves the sensitivity by 13\%. This translates in 30\% reduction in observing time which is like adding a fourth telescope to the three already available. The weighted scheme for track analysis is now considered the baseline for IXPE data analysis.

IXPE data are analyzed on ground, as briefly explained in \cite{Soffitta21}. Binary telemetry files are converted into FITS standard format containing all the information provided by the Instrument and other relevant information from the spacecraft. In particular, they contain track images that are analyzed to estimate the Stokes parameters and $\alpha^{0.75}$ that is the best weight available to be used, as explained in this paper. IXPE data will be publically released to the scientific community with a format including for each event a weigjt to apply, given by $\alpha^{0.75}$.

\section{Acknowledgments}
The Italian contribution to the IXPE mission is supported by the Italian Space Agency (ASI) through the contract ASI-OHBI-2017-12-I.0, the agreements ASI-INAF-2017-12-H0 and ASI-INFN-2017.13-H0, and its Space Science Data Center (SSDC), and by the Istituto Nazionale di Astrofisica (INAF) and the Istituto Nazionale di Fisica Nucleare (INFN) in Italy.

IXPE is a NASA Astrophysics Small Explorers (SMEX) mission, managed by MSFC and overseen by the Explorers Program Office at GSFC.

\bibliography{paper_weights_bibl}{}
\bibliographystyle{aasjournal}



\end{document}